\documentclass[aip,numerical,reprint,amsmath,amssymb]{revtex4-1}

\usepackage{graphicx}
\usepackage{dcolumn}
\usepackage{bm}

\newcommand{\sbr}[1]{\left[#1\right]}
\renewcommand{\Re}{\mathrm{Re}}

\begin{document}

\preprint{Accepted for publication in Applied Physics Letters (March 2012)}

\title{Quantum Spin Holography with Surface State Electrons}

\author{Oleg O. Brovko}
\affiliation{Max-Planck-Institut f\"{u}r Mikrostrukturphysik, Weinberg 2, D06120 Halle, Germany}
\email{obrovko@mpi-halle.de}
\author{Valeri S. Stepanyuk}
\affiliation{Max-Planck-Institut f\"{u}r Mikrostrukturphysik, Weinberg 2, D06120 Halle, Germany}

\date{\today}

\begin{abstract}
    In a recent paper Moon and coworkers [C.R.~Moon \textit{et~al.}, Nature Nanotechnology 4, 167 (2009)] have shown that the single-atom limit for information storage density can be overcome by using the coherence of electrons in a two-dimensional electron gas to produce quantum holograms comprised of individually manipulated molecules projecting an electronic pattern onto a portion of a surface. We propose to further extend the concept by introducing quantum spin holography -- a version of quantum holographic encoding allowing to store the information in two spin channels independently.
\end{abstract}


\maketitle


    In the era of ever increasing information storage density the physical entity setting the limit to the progress has quickly been reduced to the size of a single atom. The concept of representing a single bit of information by a single atom or molecule has been conceived with the ability to manipulate individual atoms on a surface with the tip of a scanning tunneling microscope. \cite{Eigler1990} Two decades later it has also been shown that it is possible to store information in the state of a single free-standing atom trapped in an optical cavity \cite{Specht2011} and that constructing truly atomic logic elements is not any more beyond the grasp of state-of-the-art experimental techniques \cite{Heinrich2002,*Khajetoorians2011,*Heinrich2011} And despite the fact that achieving information storage densities close to the single-atom limit in real-world devices is still far from being feasible, the search for an information carrier which would break the limit set by the finite spacing between single atoms in solid state systems has not been abandoned.

\begin{figure}[b]
		\includegraphics[scale=0.85]{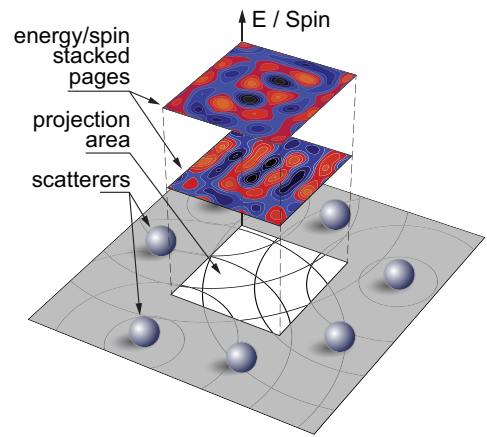}
        \caption{\label{fig:01}(color online) Principles of quantum holographic encoding in a two-dimensional electron gas. Inspired by Ref.~\onlinecite{Moon2009}.}
    \end{figure}

    The main requirement that such a carrier has to fulfill is that it's characteristic lateral dimensions should be smaller than those of a single atomic unit. In a recent paper Moon and coworkers have proposed that information can be stored in a fermionic state of a coherent two-dimensional electron gas. \cite{Moon2009,*Heller2009} They have dubbed the proposed concept \emph{quantum holographic encoding}, drawing parallels between the free space optical waves and electron wavefuctions. They use atomic manipulation \cite{Eigler1990} to construct \emph{molecular holograms} of carbon monoxide (CO) molecules on the (111) surface of a Cu crystal (Fig.~\ref{fig:01}) hosting a two-dimensional quasi-free electron gas -- a Shockley type surface state (SS). \cite{Oka2010,*Negulyaev2009,*Shockley1939}
    Electron waves scatter at local potentials introduced by the molecules and interfere (owing to inherent long range coherence \cite{Silly2004}) in a designated area of the surface (white area in Fig.~\ref{fig:01}) to form an electron density of states (LDOS) pattern representing an information page. \cite{Moon2009} This page can then be read out with a density sensitive scanning probe technique, such as the scanning tunneling microscopy (STM). There are several major advantages of such information encoding procedure: the information is projected onto a medium (i) free of lattice constraints and (ii) native to the system itself. This means that the size of a single bit is limited neither by the host surface lattice nor by the limitations of a readout technique (f.e. by lensing or collimation of an external readout beam \cite{Moon2009}).

    Furthermore, owing to the parabolic dispersion of the SS, the interference patters formed by electrons with different energies are different. Using this property Moon \emph{et al.} were able to project the hologram not only in two spatial degrees of freedom but also in the energy dimension (Fig.~\ref{fig:01}), a concept very similar to optical volumetric holography. \cite{Heanue1994,Moon2009}

    It is, however, conceivable, that in the same way as Moon and coworkers \cite{Moon2009} have utilized the energy dispersion, other intrinsic properties of electrons can act as a new dimension for information storage, pushing further back the information density limit. In the present letter we propose to use the spin of electrons as such additional dimension. We show that if the molecules or atoms used in the construction of a molecular hologram have pronounced magnetic properties or the surface state is inherently spin-polarized the scattering of surface state electrons becomes spin-dependent, allowing one to store different information pages in different spin channels of a collinear uniaxial system. As an example we discuss Cu(111) surface decorated with bilayer cobalt (Co) islands capped with Cu thus obtaining a system with spin-polarized surface states. \cite{Diekhoner2003} With its help we demonstrate the possibility of simultaneously encoding different information pages with electrons of the same energy but opposite spins as sketched in Fig.~\ref{fig:01}.

    In order to have an effective means to treat the quantum holographic encoding theoretically we adopt the multiple scattering formalism, which has in the past proven to be a uniquely suitable tool for the description of surface state scattering by impurities. \cite{Heller1994} In terms of complex amplitudes the process of scattering at a single potential modifies the amplitude of electrons as
    \begin{equation}a_{out}=a_{in}\cdot\frac{e^{2i\eta}-1}{2i}\;,\end{equation}
    where $a_{in}$ is the incoming amplitude and $\eta$ is the complex phase shift. The latter can be decomposed into real and imaginary parts such that $e^{2i\eta}=\alpha \exp(2i\delta)$, where the real part $\delta$ describes the scattering phase and $\alpha$ describes the attenuation of the amplitude due to various processes (e.g. adsorption or, in particular case of an atom on a surface, scattering into bulk states). Parameters $\alpha$ and $\delta$ are determined by the shape of the scattering potential and, in turn, uniquely define the process of scattering. The propagation of an electron across the surface can, to the first order, be considered as a free electron propagation \cite{Heller1994}:
    \begin{equation}a=a_0 \cdot \sqrt{\frac{2}{\pi k \Delta r}}\cdot e^{ik\Delta r}\;,\end{equation}
    where $k$ is the wave length and $\Delta r$ is the propagation distance.
    The local density of states (LDOS) at a certain point $r$ induced by a single atom located at $r_i$ is then determined by the interference of outgoing (in the spirit of the Huygens principle) and scattered amplitudes at that particular point. Considering the outgoing amplitude to be an order zero bessel function (being equal to 1 at the origin), the LDOS is then proportional to:
    \begin{equation}\label{eq:singscat}LDOS(r,k) \varpropto \Re[a] = \Re\sbr{\frac{\alpha e^{2i\delta}-1}{2i}\frac{e^{2ik(r_i-r)}}{r_i-r}}.\end{equation}
    The generalization to the multiple scattering case of $n$ potentials is straightforward \cite{Heller1994}:
    \begin{equation}\label{eq:multscat}a_{MS}(r,k) = \bf {a_{o}} [1-\bf A]^{-1} \bf {a_{s}}\;,\end{equation}
    where $\bf {a_{o}}$ and $\bf {a_{s}}$ are vectors of length $n$ describing the propagation of the outgoing and scattered waves to and from scatterers $i=1..n$, and $\bf A$ is an $n\times n$ matrix describing electron propagation between individual scatterers. The spatial LDOS distribution produced by an array of scatterers in a surface state is thus determined by 3 parameters: the $k$-vector of electrons (tied to their energy by the dispersion relation) and the parameters $\alpha$ and $\delta$ describing the scattering potentials. Moon \emph{et al.} have made use of the energy dispersion of the surface state ($k=k(E)$) to open the energy dimension for quantum holographic encoding. \cite{Moon2009} We, however would like to utilize the fact, that some or all of the three above named parameters might be spin dependent.

\begin{figure*}
		\includegraphics[scale=1.0]{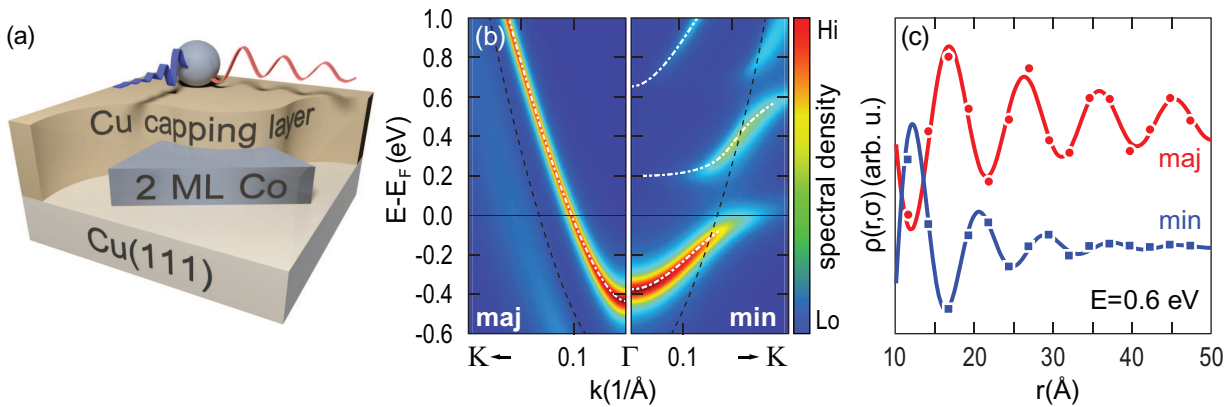}
        \caption{\label{fig:02}(color) (a) Sketch of the system: bilayer Co island on Cu(111) covered with two monolayers of Cu. (b) Spectral density map of majority (left) and minority (right) electrons (along the $\mathrm{K}-\Gamma-\mathrm{K}$ direction of the surface Brillouin zone) some $5~\mathrm{\AA}$ above the system shown in (a). (c) Spin resolved Friedel oscillations at $0.6~\mathrm{eV}$ around a singe Cu adatom on Cu/Co/Cu(111). The data points represent the KKR calculation, the solid line is a fit of Eq.~\ref{eq:singscat} to the data (modified for minority electrons to account for a faster decay).}
    \end{figure*}

    Two possibilities to achieve such spin-dependence spring to mind. One would be to use magnetic atoms or molecules for the construction of the hologram. The scattering potentials that they present for the surface state electrons shall then be different for electrons with differently oriented spins. In a uniaxial system one could say that $\alpha=\alpha(\sigma)$ and $\delta=\delta(\sigma)$ become dependent on the spin $\sigma$ of the electron (either $\uparrow$ or $\downarrow$ in a uniaxial system), so that the same arrangement of scatterers would produce different LDOS patterns in different spin channels. In realization of such a scenario care should be taken to ensure the uniaxial character of the spins' alignment, their thermal stability and the absence of Kondo screening of the atomic or molecular moments. The first two and to some extent the third criterion could be satisfied, f.e., if one uses a substrate with substantial crystalline anisotropy and/or couples the spins of the adatoms or molecules constituting the hologram to an external magnetic field \cite{Meier2008} or an underlying subsurface magnetic layer. \cite{Brovko2008prl,Uchihashi2008} To avoid the Kondo screening one could also concentrate on a system with a Kondo temperature below the experimental conditions. We, however, concentrate on another possibility of making electron scattering spin-dependent. Instead of using magnetic statterers, we switch to a system with an inherently spin-polarized surface state -- Co nanoislands or multilayers on Cu(111). \cite{Oka2010} This system is well studied both experimentally and theoretically \cite{Diekhoner2003,Oka2010,Oka2011} and provides us exactly with what we need -- the spin-dependence of scattering parameters $\alpha$ and $\delta$ as well as the spin-dependence of the electron wave vector $k$. However, Co nanoislands do themselves scatter the surface state electrons \cite{Oka2010} and would thus slightly interfere with the encoding of the hologram. To avoid that, and at the same time protect the system from additional contamination and intermixing, the Cu surface with islands grown on top can be covered with a few additional layers of copper thus sealing the island within the surface and reducing their effect on the scattering of surface state electrons while still retaining the spin-polarized character of the surface state. Without limiting the generality we shall, in the following, concentrate on a particular example of such a system -- bilayer cobalt nanoislands adsorbed on a Cu(111) surface and capped with two additional layers of copper [Cu/Co/Cu(111)]. A sketch of the system is given in Fig.~\ref{fig:02}(a). We treat the system theoretically with a first-principles code based on Korringa-Kohn-Rostoker (KKR) Green's function method (in the framework of the density functional theory). This method is described in detail in numerous publications \cite{Wildberger1995,Zeller1995,*Negulyaev2008,*Negulyaev2009} and has been shown to be well suited for describing the kind of systems we are about to address. \cite{Oka2010}


    The calculated spectral density maps of electrons some $5~\mathrm{\AA}$ above the surface of the system described above is presented in Fig.~\ref{fig:02}(b) for majority (left) and minority (right) electrons. It is apparent that the surface band structure of the system is spin-polarized. While majority electrons show a parabolic dispersion band (white dash-dotted line in the left panel) residing in the projected copper bulk band gap (black dashed lines), the minority electrons (right panel) form a dispersive band below the Fermi energy with a relatively high effective mass and have a number of additional dispersive features in the energy ranges of $0.2-0.6~\mathrm{eV}$ (arising from the hybridization of surface $sp$ electrons with a $d$-level of Co) and above $0.7~\mathrm{eV}$ (the analogue of the majority parabolic surface state band pushed beyond the Fermi level). It is worth noting, that those bands cross the boundary of the projected bulk band gap, which in normal circumstances would lead to a formation of a fairly broad resonance. Here, however, the bands remain largely unbroadened, which means that the layered structure topologically protects the surface state from scattering into the bulk. This protection is, however, not complete as shall be seen later.

    Having at our disposal a spin-polarized surface state we can now attempt to encode a hologram using electrons of a certain energy but with different spin orientations to store different information pages. To do so, we need to extract the scattering parameters ($\alpha$, $\delta$ and $k$). In the selection of the electron energy for holographic encoding we rely on two factors: high $k$-vector to reduce the characteristic electron wavelength and thus maximize the potential information density and possibly small deviation from the Fermi energy in order to facilitate experimental realization. Based on that we select an energy of $0.6~\mathrm{eV}$ above the Fermi energy, which corresponds to the upper end of the dispersive minority band in Fig.~\ref{fig:02}(b). Parameters $\alpha$, $\delta$ and $k$ are then obtained for the given energy by fitting Eq.~\ref{eq:singscat} to a radial LDOS distribution (Friedel oscillations) produced by a single scatterer on a surface. Such a distribution for a single Cu adatom on Cu/Co/Cu(111) is given in Fig.~\ref{fig:02}(c) for majority (red circles) and minority (blue rectangles) electrons. It is apparent that the Friedel oscillations of majority and minority electrons are substantially different. Not only are they defined by different scattering parameters, but they also seem to have different decay ratios. While the majority curve can be well fitted with Eq.~\ref{eq:singscat}, yielding the red solid line, the minority curve seems to decay exponentially rather than follow the $1/r$ rule \footnote{In fact, since the decay path cannot be unambiguously identified from our calculations, the correct analytic decay rate cannot be guaranteed. The exponential factor, however, yields much more accurate fitting results, than $1/r^n$ power law.}. This is the consequence of the upper part of the minority band (at $0.6~\mathrm{eV}$), with which the scattered electrons can be associated, lying outside the projected copper bulk band gap. While the Co bilayer island and the capping Cu bilayer seem to largely prevent the surface state electrons from being scattered into the bulk, the probability of such a scattering is non-zero and thus the coherence length of surface state electrons is reduced causing the exponential, rather than $1/r$, decay of Friedel oscillations. The blue solid curve represents a fit of the data to Eq.~\ref{eq:singscat} with an additional exponential factor added. Nevertheless ,the conditions for the spin-selective encoding are satisfied and we can assume that encoding different information pages with electrons of different spin character is possible. Following the notation of Moon and coworkers such encoding could be called the ``quantum spin holography'' (QSH).

    \begin{figure*}
		\includegraphics[scale=1.0]{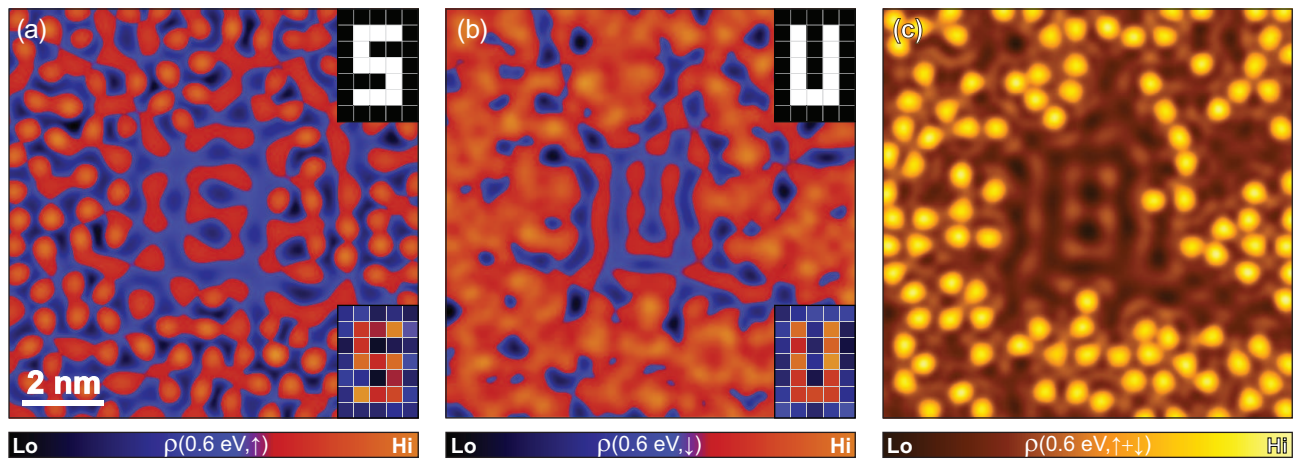}
        \caption{\label{fig:03}(color) Quantum spin hologram, encoding two information pages in different spin channels. Bit maps $5\times 7$ pixels in likeness of letters ``S'' (a) and ``U'' (b) encoded with majority and minority electrons, respectively. The insets show target (top) and resulting (bottom) information pages. On the right (c) the total LDOS distribution $\rho(tot)=\rho(E_F,\uparrow)+\rho(E_F,\downarrow)$ is shown.}
    \end{figure*}

    To prove the possibility of QSH we will try to encode the patterns used by Moon \emph{et al.} in the ``volumetric quantum holography'' section of their paper \cite{Moon2009} -- namely $5\times 7$ bit pages in likeness of letters ``S'' and ``U''. The procedure of encoding the quantum spin hologram is virtually identical to the one described by Moon \emph{et al.} for the volumetric quantum holography: one has to find a scatterer distribution producing the desired pattern in the target area of the surface (Fig.~\ref{fig:01}). The pattern produced by a given atom distribution (atomic hologram) was calculated using the multiple scattering formalism as formulated in Eq.~\ref{eq:multscat}. The corresponding parameters $\alpha$, $\delta$ and $k$ were obtained by fitting as described above. The quality of the pattern created by a particular distribution was assessed by calculating the standard Pearson correlation \cite{Rodgers1988} between the calculated LDOS pattern [digitized on a dense real space mesh and averaged locally to obtain the $5\times 7$ bit pattern, see lower insets in Figs.~\ref{fig:03} panels (a) and (b)] and the target bit map [upper insets in Fig.~\ref{fig:03}(a) and (b)]. The correlation was then maximized over the whole set of possible atomic/molecular configurations to optimize the hologram. While theoretical annealing was effectively utilized by Moon and coworkers \cite{Moon2009} to predict optimal molecular positions for volumetric quantum holography, we chose make use of a genetic algorithm technique -- a search heuristic that mimics the process of natural evolution \cite{Dudiy2006,Forrest1993} to enhance the theoretical annealing method. Genetic algorithms allow faster convergence and increase the chance of avoiding local minima in the optimization of the pattern-target correlation. In optimizing the atomic configuration of the molecular hologram, the following experimental necessity was taken into account: the atoms were prohibited to form dimers as those are virtually impossible to control with an STM tip.

    The procedure described above applied to an area of Cu/Co/Cu(111) surface approximately $10\times 10~\mathrm{nm}$ with a target area of $2\times3~\mathrm{nm}$ yielded the holographic patterns shown in Fig.~\ref{fig:03}(a) for majority and in Fig.~\ref{fig:03}(b) for minority electrons at the Fermi level. The optimized positions of atoms forming the hologram can be deduced from the map of the total density of states in Fig.~\ref{fig:03}(c). It is apparent that while the total LDOS map [Fig.~\ref{fig:03}(c)] contains a seemingly random pattern of spots the majority and minority [Figs.~\ref{fig:03}(a) and \ref{fig:03}(b)] LDOS patterns closely resemble (within the projection area in the center) the targeted information pages (upper insets). To quantify the information contained in the hologram we apply the procedure used by Moon \emph{et al.} and dividing the projection area into a grid of $5\times 7$ squares average the LDOS within each square. The obtained values can then be compared to a predefined threshold resulting in 35 bits of encoded information). The result of averaging of the holograms presented in panels (a) and (b) of Fig.~\ref{fig:03} is presented in the lower insets in the corresponding panels.


    In summary, we have demonstrated the feasibility of quantum spin holography, \textit{i.e.} information can be stored in the local density of states patterns created independently by majority and minority electrons scattered at an ensemble of magnetic atoms. Introducing magnetism into the system thus theoretically allows one to double the information density as compared to the volumetric quantum holographic encoding. \cite{Moon2009} We are also rather optimistic about the possibility of the experimental realization of the proposed concept using a spin-polarized scanning tunneling microscope for construction and readout of quantum spin holograms.

    \bibliography{qsholo}
\end{document}